\documentclass{iaus}
\usepackage{graphicx}
\newcommand{\tal}{\it et al. \rm}

\title[Boxy/peanut and discy bulges : formation, evolution and properties] %% give here short title %%
{Boxy/peanut and discy bulges : formation, evolution and properties}

\author[E. Athanassoula]   %% give here short author list %%
{E. Athanassoula}

\affiliation{Laboratoire d'Astrophysique de Marseille, Observatoire
  Astronomique de Marseille Provence, 2~place Le Verrier, 13248
  Marseille cedex 04, France\\
}

\pubyear{2008}
\volume{245}  %% insert here IAU Symposium No.
\pagerange{1--10}
\date{?? and in revised form ??}
\jname{Proceedings Title IAU Symposium}
\editors{A.C. Editor, B.D. Editor \& C.E. Editor, eds.}

\begin{document}

\maketitle

\begin{abstract}
The class `bulges' contains objects with very different formation and
evolution paths and very different properties. I review two
types of `bulges', the boxy/peanut bulges (B/Ps) and the discy bulges. The
former are {\it parts} of bars seen edge-on, have their origin in  
vertical instabilities of the disc and are somewhat shorter
in extent than bars. Their stellar population is similar to
that of the inner part of the disc from which they formed. 
Discy bulges have a disc-like outline, i.e., seen face-on they are
circular or oval and seen edge-on they are thin. Their extent is of
the order of 5 times smaller than that of the boxy/peanut bulges. They
form from the inflow of 
mainly gaseous material to the centre of the galaxy and from
subsequent star formation. They thus 
contain a lot of young stars and gas. Bulges of different types often
coexist in the same galaxy. I review the main known results on these two
types of bulges and present new simulation results.  

B/Ps form about 1Gyr after the bar, via a vertical buckling. At that
time the bar strength decreases, its inner part becomes thicker --
forming the peanut or boxy shape -- and the ratio
$\sigma_z^2/\sigma_r^2$ increases. A second buckling
episode is seen in simulations with strong bars, also accompanied by a
thickening of the peanut and a weakening of the bar. 
The properties of the B/Ps correlate strongly with those of the bar:
stronger bars have stronger peanuts, a more flat-topped
vertical density distribution and have experienced more bucklings.

I also present simulations of disc galaxy formation, which include the
formation of a discy bulge. Decomposition of their radial
density profile into an exponential disc and a S\'ersic bulge gives
realistic values for the disc and bulge scale-lengths and mass ratios,
and a S\'ersic shape index of the order of 1.     

It is thus clear that classical bulges, B/P bulges and discy bulges
are three distinct classes of objects and that lumping them together
can lead to confusion. To avoid this, the two latter could be 
called B/P features and inner discs, respectively.
 
\end{abstract}

\firstsection % if your document starts with a section,
\section{Introduction}

What is a bulge? Three different definitions have been used so far,
based on morphology, photometry, or kinematics, respectively. 
According to the morphological definition, a bulge is the component of a
disc galaxy that swells out of the central part of a disc viewed
edge-on. Based on photometry, a bulge is the extra light in the
central part of the galaxy, over and above the exponential profile
fitting the remaining (non 
central) part of the disc. The third definition is based on 
kinematics, and in particular on the value of $V/\sigma$, or, more
specifically, on the location of the object on the ($V/\sigma$,
ellipticity) diagram (often referred to as the Binney diagram
(\cite[Binney 1978]{Binney78}, \cite[2005]{Binney05})). These three
definitions are compared and discussed in \cite{AthMV07}.

The lack of a single, clear-cut definition of a bulge, although
historically understandable, has led to considerable confusion and to the fact
that bulges are an inhomogeneous class of objects. For this reason, 
Kormendy (\cite[1993]{Kormendy93}; see also
\cite[Kormendy \& Kennicutt 2004]{KormendyKennicutt04}, hereafter
KK04) distinguished classical bulges from pseudo-bulges. 
However, pseudo-bulges by themselves are also 
an inhomogeneous class of objects, as argued by \cite[Athanassoula
  (2005a, hereafter A05)]{A05bulges}, 
who distinguishes three types of objects which are, according to the
above definitions, classified as bulges. 
{\bf Classical bulges} are formed by gravitational collapse or 
hierarchical merging of smaller objects and corresponding dissipative gas
processes. Their morphological, photometrical and kinematical properties 
are similar to those of ellipticals. They are discussed extensively in
other papers in these proceedings and are not the subject of this
review. The two other types of bulges are {\bf boxy/peanut bulges}
(B/P), and {\bf discy bulges}, which will be discussed here.
As stressed in A05,
different types of bulges often co-exist and it is 
possible to find all three types of bulges in the same simulation,
or in the same galaxy. 

\section{Boxy/peanut bulges}\label{sec:bpbulges}

Viewed edge-on, disc galaxies often have a central component which
swells out of the disc and whose outline is not elliptical, but has a
boxy, or peanut, or even `X' shape. Due to the morphological definition
of a bulge, such components have been called bulges, or, more
specifically, boxy/peanut bulges (B/Ps). 

The formation of B/Ps has been witnessed in a large number of
numerical simulations (\cite[Combes \& Sanders 1981]{CombesSanders81};
\cite[Combes \tal 1990]{CDFP90}; \cite[Raha \tal 1991]{RahaSJK91};
\cite[Athanassoula \& Misiriotis 2002, hereafter AM02]{AM02};
\cite[Athanassoula 2003, hereafter A03]{A03}; \cite[A05]{A05bulges};
\cite[O'Neil \& Dubinski 
  2003]{ONeilDubinski03}; \cite[Debattista \tal
  2004]{DebattistaCMM04}; \cite[Martinez-Valpuesta \& Shlosman
  2004]{MartinezVShlosman04}; \cite[Debattista \tal
  2006]{DebattistaCMMWQ06}; \cite[Martinez-Valpuesta, Shlosman \&
  Heller 2006]{MartinezVSH06}, etc). It is linked to the vertical
instability of 
parts of the main family of periodic orbits constituting the bar,
widely known as the x$_1$ family (\cite[Binney 1981]{Binney81};
\cite[Pfenniger 1984]{Pfenniger84};
  \cite[Skokos, Patsis \&
Athanassoula 2002]{SkokosPA02}; \cite[Patsis, Skokos \& Athanassoula
      2002]{PatsisSA02}). The stability of the x$_1$ family can be  
followed from the corresponding stability diagram (see e.g. figures 3
and 4 of \cite[Skokos \tal 2002]{SkokosPA02})  
which shows that, at the positions where the x$_1$ becomes unstable, other
families bifurcate. These are linked to the $n:1$ vertical resonances
and extend well outside the disc equatorial plane. As shown by Patsis
\tal (2002), 
some of them are very good building blocks for the formation of
peanuts, because they are stable and because their orbits have the
right shape, extent and location. Studies of these orbits reproduced many
of the B/P properties and helped explaining crucial aspects of B/P
formation and evolution. For example, an analysis of the orbital
families that constitute peanuts predicts that B/Ps should
be shorter than bars. This is indeed found to be the case both in $N$-body
simulations and in real galaxies (\cite[L\"utticke, Dettmar \&
  Pohlen 2000]{LuttickeDP00}; \cite[A05]{A05}; \cite[Athanassoula \&
  Beaton 2006]{AthanassoulaBeaton05}).  

\subsection{Time evolution}
\label{subsec:tevol}

\begin{figure}[!h]
%\vspace{2.9in}
\centering
\includegraphics[height=4in]{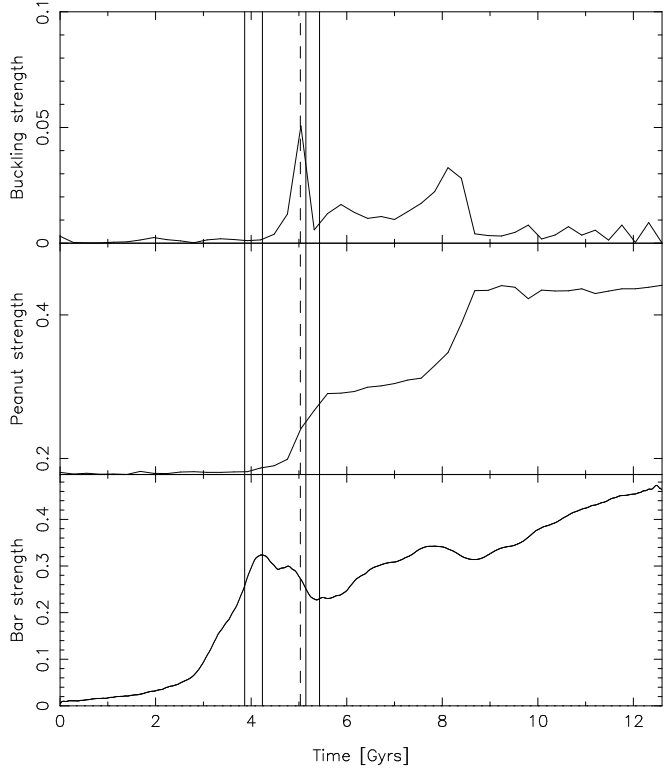}
\caption[]{ Time evolution of three peanut-, or bar-related quantities,
  namely the buckling strength (i.e. the vertical asymmetry; upper
  panel), the peanut strength (i.e. its vertical extent; middle panel)
  and the bar strength (lower panel). The solid 
  vertical lines mark characteristic times linked to bar
  formation and evolution. From left to right, these are the bar
  formation time, the maximum amplitude time, the bar decay time and
  the bar minimum amplitude time (see text). The vertical dashed line
  marks the time of the buckling.}
\label{fig:tevl}
\end{figure}

The time evolution of the bar, of the buckling and of the peanut strengths 
are plotted in Fig.~\ref{fig:tevl} for a simulation which develops a
strong bar. The time is given in Gyrs, using the calibration proposed
in \cite[AM02]{AM02}.
The initially unbarred disc forms a bar roughly between times 3 and
4 Gyrs (lower panel). I define as bar formation time the time at which
the bar-growth 
is maximum (i.e. when the slope of the bar strength as a function of
time is maximum) and indicate it by the first vertical line in
Fig.~\ref{fig:tevl}. The bar strength reaches a maximum at a time
indicated by the second vertical line, and then decreases considerably over
$\sim$ 1 Gyr. The time at which the bar amplitude decrease is maximum is
given by the third vertical line. Subsequently, the bar strength reaches a
minimum, at a time shown by the fourth vertical line, and then starts
increasing again at a rate much slower than that during bar
formation. 

The upper panel shows the buckling strength, i.e. the vertical 
asymmetry as a function of time. The disc is
vertically symmetric before and during bar formation and the first
indications of asymmetry occur only after the bar amplitude has
reached a maximum. The asymmetry then grows very abruptly to a
strong, clear peak and then drops equally abruptly. The time of the
buckling (dashed vertical line) is given by the peak of this curve and
is very clearly 
defined. It is important to note that, to within the measuring errors,
it coincides with the 
time of bar decay (third vertical line). This is not accidental. I
verified it for a very large number of simulations and thus can
establish the link between the buckling episode and the decay of the bar
strength (Raha \tal 1991; Martinez-Valpuesta \& Shlosman
2004). 

The middle panel shows the strength of the peanut, i.e. its
vertical extent, again as a
function of time. This quantity grows abruptly after the bar
has reached its maximum amplitude and during the time of the buckling.
This abrupt growth is followed by a much slower increase over a longer
period of time. Taken together, the three panels of
Fig.~\ref{fig:tevl} show that the bar forms 
vertically thin, and only after it has reached a maximum strength does the
buckling phase occur. During the buckling time the bar strength
decreases significantly, while the the B/P strength increases. The time
intervals during which bar 
formation, peanut formation, or buckling occur are all three rather
short, of the order of a Gyr, and they are followed by a
longer stretch of time during which the bar and B/P evolve much
slower. This later evolution is often referred to as secular evolution.

This particular simulation has a second, weaker buckling episode
shortly after 8 Gyrs. This occurs very often in simulations developing strong
bars and was discussed first by \cite{A05gdansk} and
\cite[Martinez-Valpuesta \tal (2006)]{MartinezVSH}. It is seen clearly
in all three panels and has characteristics similar to those of the
first buckling. 
 
\subsection{Peanut formation and collective effects}

As already discussed, orbital structure theory explains B/P formation
by the vertical 
instabilities of the main family of bar-supporting periodic orbits. 
An alternative approach explains the buckling and the peanut formation as
due to the bending, or fire-hose, instability, studied analytically in
the linear regime (\cite[Toomre 1966]{Toomre66}; \cite[Araki
  1985]{Araki85}). These studies assign  
a critical value to the ratio $R_{\sigma}=\sigma_z^2/\sigma_r^2$ igniting
the onset of the instability, which is around 0.1. A number of
simulations, however, have shown that the vertical instability sets in 
at much larger values of $R_{\sigma}$ (e.g. \cite[Merritt \& Sellwood
  1994]{MerrittSellwood94}; 
\cite[Sotnikova \& Rodionov 2003]{SotnikovaRodionov03}).  

\begin{figure}[!h]
\centering
\includegraphics[height=3.3in,angle=-90]{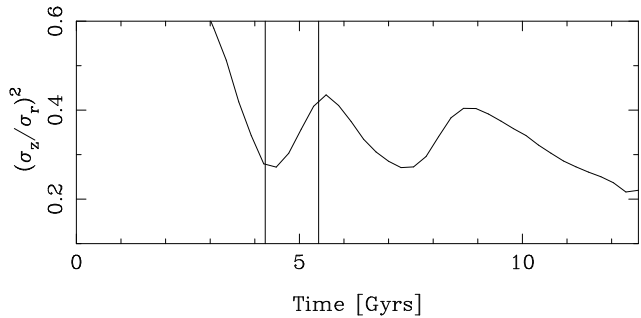}
\caption{Time evolution of the ratio $R_\sigma =
  \sigma_z^2/\sigma_r^2$. The thin 
  vertical lines mark two characteristic times linked to bar
  formation and evolution. From left to right, these are
  the bar maximum and minimum amplitude times, 
  corresponding to the first buckling episode (see Sect. \ref{subsec:tevol} and
  Fig.~\ref{fig:tevl}).  
} 
\label{fig:sigrat} 
\end{figure}

To test this hypothesis, I calculate the radial and $z$
components of the disc velocity dispersion  
as a function of radius (averaging over azimuth and height). I then
find the minimum value of their ratio $R_{\sigma}$ and
plot its time evolution in Fig.~\ref{fig:sigrat}. 
The thin vertical lines mark two characteristic times linked
to bar formation and evolution -- namely the bar
maximum and minimum amplitude times -- as found from Fig.~\ref{fig:tevl}.
Their location is clearly linked to changes in behaviour 
of $R_{\sigma}$. This, however, does not
necessarily imply that the changes in $R_{\sigma}$ are the cause of
the buckling, but can also be seen 
as its consequence. Indeed, as the bar forms $\sigma_r$ increases
drastically, so that $R_{\sigma}$ decreases. Then the bar
amplitude reaches its maximum and starts decreasing, while the peanut
starts forming. During this time, $\sigma_r$ decreases while
$\sigma_z$ increases. As a result, the ratio $R_{\sigma}$ reaches
a minimum when the bar amplitude is maximum and then increases again,
as is indeed seen in Fig.~\ref{fig:sigrat}. Then the
bar amplitude reaches a minimum, which corresponds to a minimum of
$\sigma_r$ and therefore to a maximum of $R_{\sigma}$. This is
followed by a slower decrease of $R_{\sigma}$, which is stopped by  
the second buckling 
episode. The value of $R_{\sigma}$ at which this instability sets in
is much less extreme than that predicted by the above mentioned
analytical works, but is in good
agreement with other $N$-body simulations. 

More work is necessary before we fully understand the respective roles
of the orbital structure results and of the velocity anisotropy
effects on the formation and evolution of B/P structures. Both 
explain part of the story, but many aspects of their interplay are
still unclear. Orbital structure results tell us whether the 
appropriate building blocks are available, or not, and this is
essential, since the lack of such building blocks
prohibits the formation of a given structure. Furthermore, studies of
the properties of the building-block orbits are essential for
understanding the properties of the B/P structures. 
Orbital structure theory, however, can not tell us how much matter
is trapped around a 
given orbit or family. Furthermore, it is necessary to group all these
building blocks into one coherent, self-consistent  unit and here
collective effects are essential. Like orbital structure, they also can 
set limits on the formation of B/P structures, as well as give
information on their properties. The respective input
from the two methods will be discussed further elsewhere.

\subsection{Comparison with observations}

The fact that B/Ps are just parts of bars seen edge-on was not
immediately accepted (see 
e.g. \cite[Kormendy 1993]{Kormendy93}). The main
arguments against it were, however, refuted in \cite[A05]{A05bulges}, with
the help of orbital structure results. Furthermore, 
considerable observational evidence argues in its favour, 
particularly detailed comparisons between observations and
simulations. 

Radial density profiles from simulations, taken along slits on, or
parallel to, the equatorial plane when the galaxy is seen edge-on
(AM02; A05) have the same characteristic signatures as
the corresponding radial light profiles (\cite[L\"utticke, Dettmar \&
  Pohlen 2000]{LuttickeDP00}; 
\cite[Bureau \tal 2006]{Bureautal06}). Similarly, density profiles
along cuts perpendicular to the 
equatorial plane (AM02; A05) show similar characteristics to analogous
observed light profiles (\cite{AronicaABBDVP} and this
volume). Further tests come from comparisons of median
filtered images of B/P systems (\cite[Bureau \tal 2006]{Bureautal06})
to similar images 
of $N$-body bars (\cite[A05]{A05bulges}). These show the same types of
characteristic features, namely four extensions out of the 
equatorial plane, which form an X-like shape, except that the four
extensions do not necessarily cross the centre. Another common feature
is maxima of the density along the equatorial plane, away from the
centre and diametrically opposite. Starting from the centre of 
the galaxy and going outwards along the equatorial plane, the
projected surface 
density first drops, then increases again to reach a local maximum and
then decreases again to the edge of the disc.   

Considerable evidence was also accumulated using kinematical
observations. Cylindrical rotation, witnessed in a number of B/P galaxies
(KK04 and references therein), is also seen in velocity fields of
strong $N$-body 
bars viewed edge-on (\cite[Combes \tal 1990]{CombesDFP90}; AM02).
Emission line spectroscopy of boxy/peanut galaxies (\cite[Kuijken \&
  Merrifield 1995]{KuijkenMerrifield95}; \cite[Bureau \&
  Freeman1999]{BureauFreeman99}; \cite[Merrifield \& Kuijken
  1999]{MerrifieldKuijken99}) shows that their major 
axis position velocity diagrams (PVDs) have a number of interesting
features, well reproduced by gas flow simulations (\cite[Athanassoula \&
Bureau 1999]{ABureau99}). In particular, the shocks along the leading
edges of the bar and the corresponding inflow 
lead to a characteristic gap in the PVDs, between the signature of the
nuclear spiral (whenever existent) and the signature of the
disc. 

Comparison of long-slit absorption line spectra (\cite[Chung \& Bureau
2004]{ChungBureau04}) of galaxies with B/Ps to similar `observations'
of $N$-body bars viewed edge-on reveals that the two have the same
characteristic features (\cite[Bureau \& Athanassoula 2005]{BureauA05}). The  
integrated light along the slit (equivalent to a major-axis light
profile) has a quasi-exponential central peak and a plateau at
intermediate radii, followed by a steep drop. The rotation curve ($V(r)$)
has a characteristic double hump. The velocity dispersion has a  
central peak, which in the centre-most part may be rather flat or may
even have a central minimum. At intermediate radii there can be a plateau
which sometimes ends on either side with a shallow maximum before
dropping steeply 
at larger radii. $h_3$ (i.e. the coefficient of the third order term
in a Gauss-Hermite expansion of the line of sight velocity
distribution) correlates with $V$ over most of the bar
length, contrary to what is expected for a fast rotating disc. All
these features are spatially correlated and are seen, more 
or less strongly, both in the observations and in the simulations
(\cite[Bureau \& Athanassoula 2005]{BureauA05}). 

\subsection{The effect of the halo}

\begin{figure}[!h]
\centering
\includegraphics[height=5.in,angle=-90]{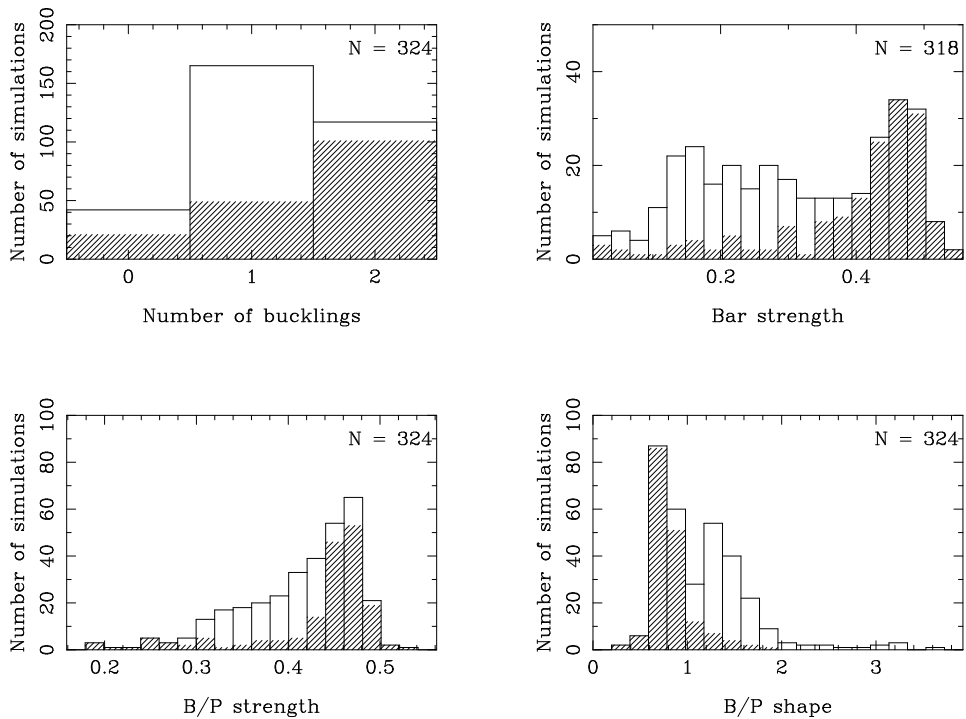}
\caption[]{Histograms of number of bucklings, bar strength, B/P strength
  and B/P shape for a sample of fully self-consistent
  high-resolution simulations. The histograms include simulations with
  halo cores of all sizes and the hatched areas includes only
  simulations with small halo cores. }
\label{fig:histo}
\end{figure}

In collaboration with Martinez-Valpuesta, I made an extensive
statistical study of a few hundred simulations which I had run for
different purposes (AM02; A03 etc). We measured the strength of
the peanut (from its thickness) and its shape (from the shape of the
density profile on cuts perpendicular to the equatorial plane) and
found that both correlate well with the bar strength. Thus, stronger
bars have stronger peanuts and more flat-topped vertical density
profiles. I
also find that the type of halo plays a major role in determining the
properties of the B/P.
Fig.~\ref{fig:histo} shows
histograms of the number of bucklings that have occurred, of the bar
strength, of the B/P strength and of the B/P shape, distinguishing
between
simulations with small halo cores and simulations with large halo
cores. The two populations 
are indeed very different. Simulations with small halo cores
have stronger bars, stronger peanuts, more
flat-topped vertical density profiles and
have experienced more bucklings than simulations with large cores. The
few simulations with small cores which have weak bars, weak B/Ps and
did not buckle have either a very hot halo or a very hot disc. 
Simulations with
cuspy haloes (not shown here) have yet weaker B/Ps and smaller number
of bucklings and will be discussed elsewhere.  

The above can be explained by the fact that the halo plays a
major role in determining the properties of the bar (AM02;
A03). \cite[Athanassoula (2002, hereafter A02)]{A02}
showed that angular momentum is primarily emitted by near-resonant
material in the bar 
region and absorbed by near-resonant material in the outer disc and,
particularly, in the halo, while A03 showed that bars grow
stronger when more angular momentum is exchanged within the
galaxy. Furthermore, as shown in AM02 and as explained in A02 and A03,
the size of the halo core strongly influences 
the bar evolution. Haloes with a small core have a lot of mass in
the inner regions and thus, provided their velocity dispersion 
is not too high, can provide substantial angular
momentum sinks and lead to considerable angular momentum
exchange between the near-resonant particles in the bar region and
the near-resonant particles in the halo. Such models 
grow strong bars (long, thin and massive) with rectangular-like
isodensities (AM02). Viewed side-on (i.e. edge-on with the
line-of-sight along the bar minor axis) they exhibit a strong peanut, or
even X-like shape. If, however, the velocity dispersion in the disc
and/or halo is too high, the angular momentum exchange is hindered and
the bar and peanut will be weak (A03). Haloes with large cores have
considerably less material in the inner parts and are thus 
capable of less angular momentum exchange.  
Bars grown in such environments are less strong, have 
elliptical-like isodensities when viewed face-on and boxy-like when
viewed side-on. All these considerations explain the results found in
Fig.~\ref{fig:histo}, namely the difference between the histograms for
simulations with small halo cores and simulations with large halo
cores. They also explain the weak bars and B/Ps found in some
simulations with small halo cores.   

\section{Disc-like bulges}\label{sec:discyb}

Disc-like bulges form from inflow of
(mainly) gas material to the centre of the galaxy and subsequent star
formation. This inflow is due to the torques exerted by a
non-axisymmetric component, usually a large-scale bar, as witnessed in
hydrodynamic simulations (e.g. \cite[Athanassoula
  1992]{A92b}; \cite[Friedli \& Benz 1993]{FriedliBenz93};  
\cite[Heller \& Shlosman 1994]{HellerShlosman94}; \cite[Wada \& Habe
  1995]{WadaHabe95}). The high density of the gas accumulated in the 
inner regions triggers very strong star formation. 
Such inflow, however, is also seen in $N$-body simulations
(\cite[AM02]{AM02}; \cite[Valenzuela \&
  Klypin 2003]{ValenzuelaKlypin03}), which represent the old stellar population. Thus, this inner
region is not only a region of increased density for the gas and the young
stars, but also for the older stellar populations. This should lead to the
formation of an inner, central component of disc-like shape, whose
extent is of the order of a kpc and which is constituted mainly of gas and
young stars, but also of older stars. This was named disc-like bulge,
or, for short, discy bulge in \cite[A05]{A05bulges} and is often
observed in disc galaxies. Due to its
disc-like shape, it often has spirals or inner bars
(KK04 and references therein). It stands out very clearly in radial
photometric profiles, whose decomposition shows it is well represented
by a S\'ersic law (\cite[S\'ersic 1968]{Sersic68}). Contrary to
classical bulges, however, it does not swell out of the galactic plane. 
This is not the only difference between disc-like and
classical bulges. Disc-like bulges have a 
S\'ersic index of the order of 1, i.e. much smaller than the
values found for classical bulges (KK04 and references therein). They 
also have different kinematics, like that of discs, a
higher fraction of young stars and a higher gas content. A lot of data
on such bulges has been collected over the last few years, but still
much work, particularly theoretical, is necessary before we fully
understand these objects. 

In order to describe adequately the formation and evolution of
disc-like bulges, simulations should include gas, star formation and
feedback, all in a realistic way. It would, furthermore, be preferable
if they started from cosmological or cosmologically-motivated initial
conditions, since the properties of pre-existing discs may influence
the properties of the disc-like bulges. I will briefly describe here
results from simulations following this outline (Athanassoula,
Heller \& Shlosman, in preparation). For information on the numerical
techniques used in these simulations and an initial discussion of some
of the results see \cite{HellerSA07a} and
\cite[(2007b)]{HellerSA07b}.   

\begin{figure}[h!]
\centering
\includegraphics[height=2.5in,angle=-90]{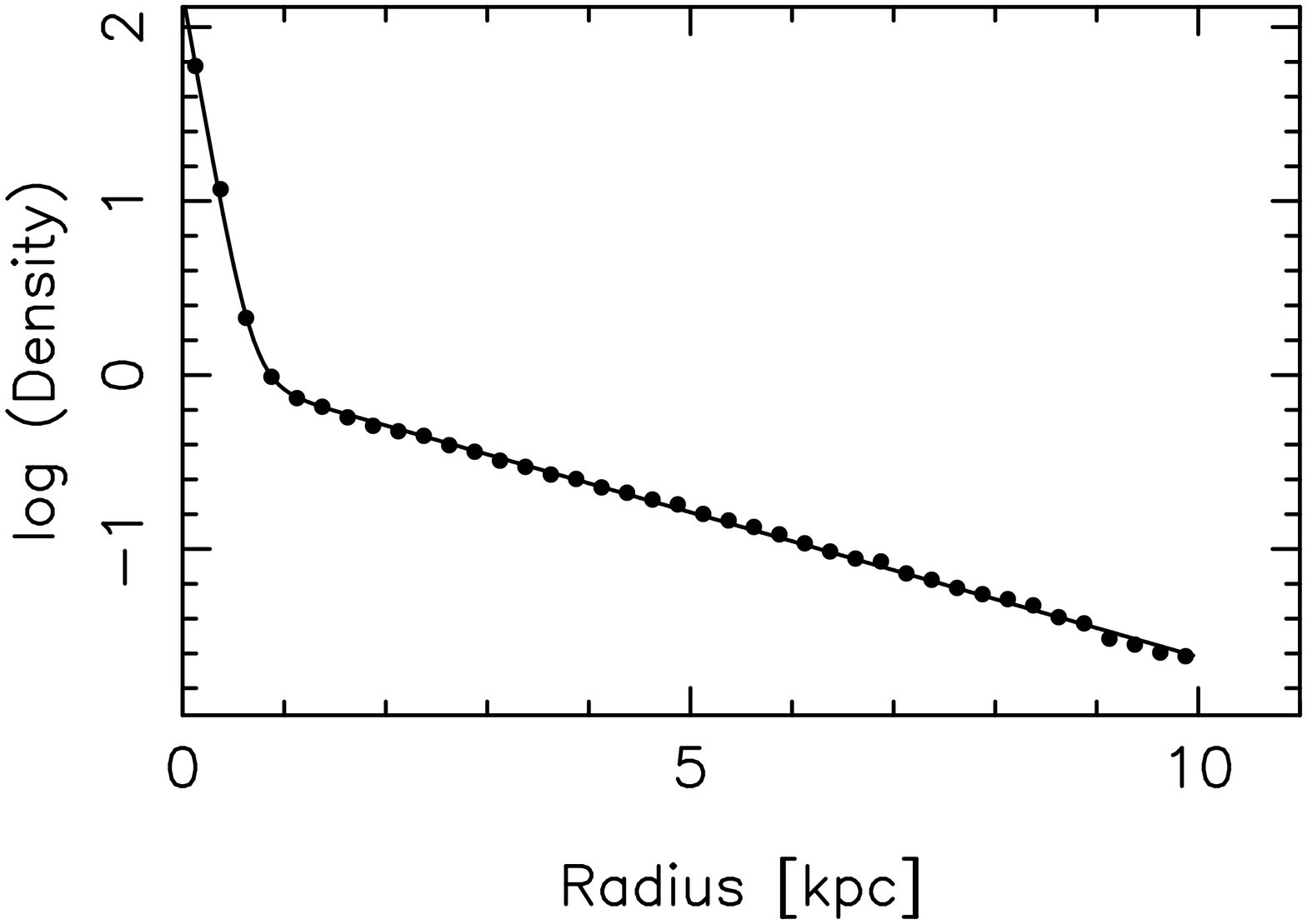}
\includegraphics[height=2.4in,angle=-90]{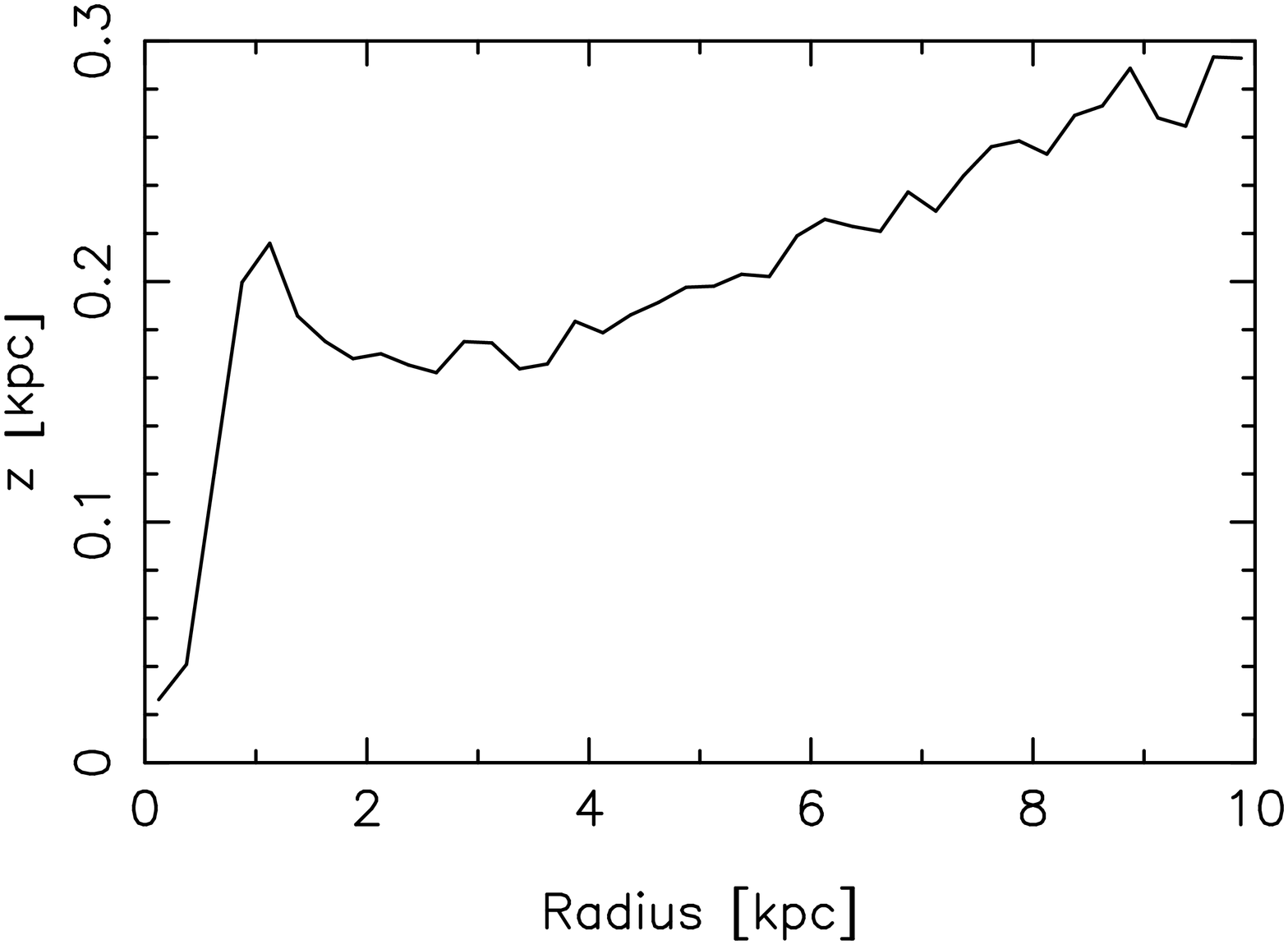}
\caption[]{Properties of a simulated disc-like bulge. Left : Radial
  projected density profile in arbitrary units. Radii are measured in
  kpc. The dots give the
  simulation results and the straight line the fit by an exponential
  disc and a S\'ersic component. Right : Measure of the vertical
  height of the material near the equatorial plane (see text), as a
  function of radius, measured in kpc. }
\label{fig:densitydiscy}
\end{figure}

Several non-axisymmetric components -- such as a triaxial halo, oval disc,
inner and outer bar -- form during these simulations. Their interactions
give very interesting dynamical phenomena (\cite[Heller et al.
 2007a]{HellerSA07a};
\cite[b]{HellerSA07b}), while they induce considerable inflow
and gaseous high density inner discs. As in the sketchy outline in the
beginning of this section, the high gas concentration in the central 
area triggers considerable star formation, resulting in a disc-like
central, high-density object, which, seen face-on, is often
somewhat oval. It has many properties similar to
those of discy-bulges. For example, it has, in many cases,
sub-structures, like an inner bar. In order to further assess the
properties of the disc galaxy formed in these simulations and to
better establish the link with disc-like bulges, I chose a
characteristic specific snapshot, i.e. a characteristic
specific simulation and time, and examine its mass distribution in
order to compare best with the observed light distribution in
galaxies. An analysis of the kinematics, together with a statistical
treatment, including other times and other 
simulations, will be given elsewhere. The radial projected surface
density profile of the snapshot under consideration  
is given in Fig.~\ref{fig:densitydiscy}, together
with a fit by an exponential disc and a S\'ersic component. Note that
the fit is excellent, all the way to the outer parts of the disc,
roughly at 10 kpc. In this example, the disc scale-length is $\sim$2.7
kpc, i.e. very realistic, while the S\'ersic index is
$\sim$1, in good agreement with observed discy bulges. 

\begin{figure}[h!]
\centering
\includegraphics[height=4in,angle=-90]{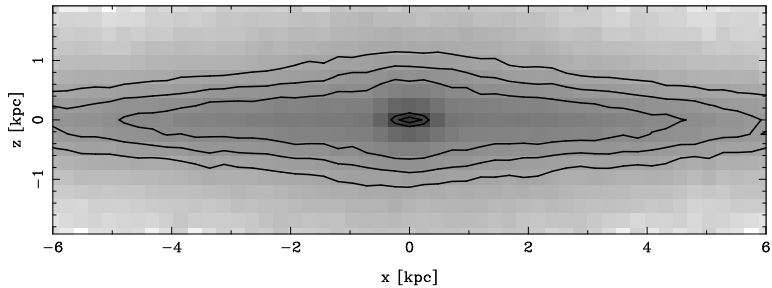}
\caption[]{Edge-on view of the stellar component of the simulation
  with a disc and a disc-like bulge. The projected density is given by  
  grey-scale and also by five isocontours whose level is picked so as
  to show best the features under consideration. }
\label{fig:edge-ondiscy}
\end{figure}

Fig.~\ref{fig:edge-ondiscy} shows the snapshot seen edge-on. The three
outer isodensity curves show clearly that the shape and aspect ratio
of the disc component is very 
realistic. The two innermost contours (within 1 kpc) reveal the
existence of a small, 
central, disc-like object, the vertical height of which we need
to quantify. Measuring the average thickness would not be useful,
since this is due to both the external big disc and the small inner
component, so I proceeded differently. I divided the
`stars' in the snapshot into circular annuli, according to their
distance from the centre and, in each annulus, sorted them as a
function of their distance from the equatorial plane ($|z|$). Since,
statistically, the `stars' in the disc-like inner component will have
smaller $|z|$ values than the ones in the outer disc, I plot in
Fig.~\ref{fig:densitydiscy} the $|z|$ component of the `star' with
rank $0.3N_{an}$, where $N_{an}$ is the total number of `stars' in the
annulus. This shows a deep minimum in the central region, as one would
expect 
due to the existence of an inner disc with a shorter vertical extent than
the outer one. It also shows that the region where the inner disc is
contributing significantly is of the order of 1 kpc, in good agreement
with the radial density profile
(Fig.~\ref{fig:densitydiscy}). Finally, the aspect ratio of the
inner and the outer discs are similar. 

To summarise, in our fiducial simulation, as well as in several
others, we witness inflow of mainly gas 
material to the central regions and strong subsequent star
formation. Thus, an inner disc is formed, composed of both stars and
gas. Its radial extent is of the order of a kpc and its vertical
extent much smaller than that of the outer disc. This disc can harbour
spiral structure, or an inner bar. Its contribution to the radial
projected density profile is well fitted by a S\'ersic law with
S\'ersic index $\sim$1. It is thus very likely that this simulation
describes correctly the formation of discy-like bulges in galaxies.

\section{Summary and discussion}\label{sec:concl}

I briefly reviewed the formation, evolution and properties of
boxy/peanut bulges and of disc-like bulges.
These two types of objects have very different formation and
evolutionary histories and very different properties. B/Ps form from
vertical instabilities and their building blocks are the 3D families
associated with the 3D bifurcations of the x$_1$ family.
Discy bulges form from the inflow of (mainly) gas material and from
the ensuing enhanced star formation. Thus B/Ps are mainly constituted
of inner disc stars, while the discy bulges have a very large contribution
from gas and young stars. Since the formation of discy bulges relies
on the gas inflow, it is expected that they will be found mainly in
late type disc galaxies, as is indeed the case. The face-on extent of
the B/Ps is of the order of five times larger than that of the discy
bulges and, seen edge-on, they extend well outside the equatorial
plane, while the discy bulges are thin. Their kinematics and their
contribution to the radial photometric profiles are different from
those of discy bulges. Thus,
one should clearly distinguish between B/Ps and discy-bulges and not
lump together them in a single category. 

Once this has become clear, one may also wish to revise the existent
nomenclature in order to avoid some of the present confusion. 
Boxy/peanut bulges 
could be called boxy/peanut features (or structures), or simply
peanuts, as proposed in \cite[A05]{A05}. This would make it clearer
that they are just a part or a feature of the bar and not an
independent entity. Similarly, discy bulges could be simply
called inner discs. Then the name `bulge' would be reserved for
classical bulges. This change, however, will also necessitate 
changing the bulge definitions described in Sect.~1. 

\begin{acknowledgments}
I thank my collaborators, A. Bosma, A. Aguerri, C. Heller,
I. Martinez-Valpuesta and I. Shlosman, for interesting and fruitful
discussions. This work was partially supported by grant 
 ANR-06-BLAN-0172 and by the Gr\"uber foundation.
\end{acknowledgments}

\end{document}